**Bright-Exciton Fine Structure Splittings in Single Perovskite Nanocrystals**

Chunyang Yin,[1] Liyang Chen,[1] Yan Lv,[1] Fengrui Hu,[1] Chun Sun,[2] William W. Yu,[2] Chunfeng Zhang,[1] Xiaoyong Wang,[1*] Yu Zhang,[2*] and Min Xiao[1,3*]

[1]*National Laboratory of Solid State Microstructures, School of Physics, and Collaborative Innovation Center of Advanced Microstructures, Nanjing University, Nanjing 210093, China*

[2]*State Key Laboratory on Integrated Optoelectronics and College of Electronic Science and Engineering, Jilin University, Changchun 130012, China*

[3]*Department of Physics, University of Arkansas, Fayetteville, Arkansas 72701, United States*

[*]Correspondence to X.W. (wxiaoyong@nju.edu.cn), Y.Z. (yuzhang@jlu.edu.cn), and M.X. (mxiao@uark.edu).

**Although both epitaxial quantum dots (QDs) and colloidal nanocrystals (NCs) are quantum-confined semiconductor nanostructures, so far they have demonstrated dramatically-different exciton fine structure splittings (FSSs) at the cryogenic temperature. The single-QD photoluminescence (PL) is dominated by the bright-exciton FSS, while it is the energy separation between bright and dark excitons that is often referred to as the FSS in a single NC. Here we show that, in single perovskite $CsPbI_3$ NCs synthesized from a colloidal approach, a bright-exciton FSS as large as hundreds of μeV can be resolved with two orthogonally- and linearly-polarized PL peaks. This PL doublet could switch to a single peak when a single $CsPbI_3$ NC is photo-charged to eliminate the electron-hole exchange interaction. The above findings have prepared an**

**efficient platform suitable for probing exciton and spin dynamics of semiconductor nanostructures at the visible-wavelength range, from which a variety of practical applications such as in entangled photon-pair source and quantum information processing can be envisioned.**

The electron-hole exchange interaction (e-h EI) is greatly enhanced in quantum-confined semiconductor nanostructures, leading to the energy-level splitting between bright- and dark-exciton states[1,2]. In semiconductor epitaxial quantum dots (QDs), the dark excitons are generally nonemissive without applying a magnetic field[3] and the bright-exciton state is further divided into two orthogonally- and linearly-polarized ones[4]. The existence of such a bright-exciton fine structure splitting (FSS) has been positively employed to generate a target wave function from coherent superposition of the two constituent states[5] and to develop a two-bit conditional quantum logic in the two-exciton configuration[6]. Meanwhile, complete elimination of this bright-exciton FSS has been actively pursued to realize a polarization-entangled photon-pair source[7-10] for fundamental tests in quantum mechanics and optics[11], as well as for practical applications in quantum communication[12-14]. Interestingly, the dark excitons in colloidal semiconductor nanocrystals (NCs) are rendered emissive due to spin-mixing interactions with several proposed sources[2,15-17] and the exciton FSS in this case usually refers to their energy separation from that of bright excitons[18,19] whose doublet states were rarely resolved. Even in the few limited reports on single CdSe NCs[20,21], the random orientation of their crystallographic axis, together with the photoluminescence (PL) blinking and spectral diffusion effects, posed severe obstacles to investigating the polarized bright-exciton FSS so that its very existence is still questionable[22].

Colloidal perovskite NCs have just emerged as a new type of semiconductor nanostructure[23,24] capable of emitting single photons without the influence of dark-exciton emission[25-27]. Moreover, the suppressions of both the PL blinking and spectral diffusion effects were successfully demonstrated in single perovskite $CsPbI_3$ NCs[28]. Here we show that the bright-exciton FSS can be easily observed in single $CsPbI_3$ NCs at the cryogenic temperature, with an energy separation as large as hundreds of μeV between the two orthogonally- and linearly-polarized states. With the laser excitation at an intermediate power, this PL doublet of neutral single exciton (X) would switch to a single peak of singly-charged single exciton ($X^-$). When the laser power is further increased, PL doublets from neutral biexciton (XX), charged biexciton ($XX^-$) and doubly-charged single exciton ($X^{2-}$) could be additionally observed. Based on the FSS values obtained from various exciton species, the isotropic and anisotropic e-h EI energies can be roughly estimated, which have provided valuable information on the fundamental electronic processes in these novel perovskite NCs.

According to the same procedure as reported previously[28], the perovskite $CsPbI_3$ NCs studied here were colloidally synthesized with a cubic size of ~9.3 nm and all the optical characterizations of single $CsPbI_3$ NCs were performed at the cryogenic temperature of ~4 K. An optical setup similar to that described in ref. 28 was utilized except that additional optical parts were added for the polarization-dependent measurements whenever necessary. The 570 nm output of a 5.6 MHz picosecond fiber laser was used as the excitation source and its output power could be adjusted to tune the number of excitons ($N$) created per pulse in a single NC.

In Fig. 1a, we plot the PL spectra collected respectively at 0 °(red), 45 °(black), and 90 ° (blue) for a representative $CsPbI_3$ NC excited at $N$ = ~0.05, where the 0 °corresponded to the angle of the polarizer set before the spectrometer to get a maximum PL intensity from the higher energy peak. The possibility that more than one NC contributed to this orthogonally-and linearly-polarized PL doublet was safely ruled out from the second-order photon correlation measurement that yielded a $g^2(\tau)$ value of ~0.05 at the zero time delay and the polarization angle of 45 °(Fig. 1b). As shown in Fig. 1c, the PL decay curves measured for the higher (red) and lower (blue) energy peaks could be both fitted with single-exponential lifetimes of ~0.93 ns and ~1.02 ns, respectively, thus confirming their bright-exciton nature of the optical emission. With a spectral resolution of ~200 μeV in the PL measurements, we plot in Fig. 1d a statistical distribution of the FSS energies measured for ~80 single $CsPbI_3$ NCs as a function of their emission energies of the higher energy peaks, from which an average value of 356 ±108 μeV could be obtained.

In our previous report[28], most of the studied $CsPbI_3$ NCs had a single PL peak with a linewidth close to ~200 μeV and only ~10% of them possessed relatively broader peaks with the PL doublets being not clearly resolved. Although the same synthesis procedure was nominally adopted, almost 80% of the total $CsPbI_3$ NCs studied here demonstrated an obvious FSS effect, implying that it is very sensitive to some trivial conditions that are too difficult to be actively controlled right now. In semiconductor epitaxial QDs, the bright-exciton FSSs of tens to hundreds of μeV are normally caused by the QD anisotropies in shape and strain[1]. The above two factors can be largely ruled out in perovskite $CsPbI_3$ NCs since they are associated with an almost symmetric shape from the TEM measurement[28], and the apparent lack of the size-dependent FSSs in Fig. 1d excludes the participation of strain-induced piezoelectric fields that should be monotonically influenced by the size/volume[29]. Since the crystal structures are intimately related to the FSSs in epitaxial QDs[30,31] and the orientational freedoms of the atomic components can substantially affect the electronic transitions in perovskite materials[32], we propose that it might be the lattice asymmetry, amenable to possible distortions in each synthesis process, that contributes to the large FSSs observed now in single $CsPbI_3$ NCs. Although the exact origins are yet to be investigated both theoretically and experimentally in future works, it does not hinder us at the current stage to explore the fundamental properties of this intriguing FSS effect in perovskite $CsPbI_3$ NCs.

These PL doublets are very robust against the measurement time, as can be seen in Fig. 1e from the time-dependent PL spectral image of a representative $CsPbI_3$ NC also excited at $N = \sim 0.05$. When the laser power was increased to make $N > \sim 0.5$, a PL doublet would switch occasionally to a red-shifted peak (see Fig. 1f and Supplementary Fig. S1a, b for a single $CsPbI_3$ NC excited at $N = \sim 0.8$). This new spectral feature was previously attributed to the optical emission from $X^-$ (ref. 28), whose total electron spin is zero so that the e-h EI vanishes to render a single PL peak without the FSS[29,33]. In previous optical studies of single perovskite NCs[27,28], both the X and $X^-$ peaks existed within a given integration time due to the extremely fast charging and de-charging events. In great contrast, the PL spectrum of $X^-$ studied here in single $CsPbI_3$ NCs could be easily separated from that of X in the time domain and even lasted for tens of minutes once $N$ was decreased to $\sim 0.1$ after it was triggered at higher laser powers (see Supplementary Fig. S1c). Because the $X^-$ state can be treated as a pure two-level system with a well-defined spin[34], its easy access and long-time stability shown here in single $CsPbI_3$ NCs highlight their great potentials in quantum technology applications such as spin storage, manipulation, and readout[35].

When the laser excitation power was further increased, optical emissions from more complex exciton species could be observed in single $CsPbI_3$ NCs, which are the hallmarks of their greatly-reduced Auger recombination processes. In Fig. 2a, we plot the time-dependent PL spectral image of a representative $CsPbI_3$ NC excited at $N = \sim 1.5$. Since the PL peak from X ($X^-$) has been well defined in Fig. 1f for this single NC excited at $N = \sim 0.8$, the accompanying one emitted at a longer wavelength can be safely attributed to the optical emission from XX (XX$^-$). We tentatively assign an additional PL peak, red-shifted only a little bit relative to that of X$^-$, to the optical emission from X$^{2-}$ due to the addition of an extra electron to the X$^-$ configuration. Corresponding to Fig. 2a, the PL spectra of X, XX, X$^-$, XX$^-$ and X$^{2-}$ are sequentially plotted in Fig. 2b-f, where the emergence of FSSs in XX and XX$^-$ is naturally expected according to previous optical studies of single epitaxial QDs[6,8-10,36,37]. In Supplementary Figs. 2 and 3, we demonstrate similar results from four more CsPbI$_3$ NCs to justify the universal appearance of all the above exciton species.

In the next, we will elaborate more on the optical properties of XX since its very existence is a prerequisite for the realization of entangled photon pairs by means of electrical[38], optical[39], magnetic[3], or piezoelectric[40] tuning of the X FSSs. In Fig. 3a, we plot the PL spectra of a representative CsPbI$_3$ NC excited at $N = $ ~1.5, with two sets of orthogonally- and linearly-polarized PL doublets originating from X and XX, respectively (see the inset for an enlarged view of the XX peaks). The higher (lower) energy X and the lower (higher) energy XX peaks have the same linear polarization because the two cascading channels, each from XX to the ground state with X being an intermediate state, should conserve the same total energy[3,38-40]. As shown in Fig. 3b, the X (XX) PL intensity has a linear (quadratic) dependence on the laser excitation power, thus confirming again the correctness of their respective origins[41]. For this specific CsPbI$_3$ NC, we also performed time-resolved measurements to obtain the single-exponential PL lifetimes of ~1.07 ns, ~0.83 ns, ~0.41 ns and ~0.36 ns for the X, X$^-$, XX and XX$^-$, respectively (Fig. 3c). Because the PL lifetime ratio of ~2.61 between X and XX is larger than the expected value of 2 (ref. 42) and the PL intensity of XX is several times lower than that of X even at the highest excitation power in our experiment (Fig. 3b), we speculate that nonradiative Auger recombination might still play a minor role in the XX decay process. From statistical measurements of ~50 single CsPbI$_3$ NCs for the energy differences between X and XX (Fig. 3d), the average binding energy of XX was calculated to be ~14.26 ± 1.53 meV, which is relatively larger than those of ~7.61 ± 1.43 meV, ~8.78 ± 1.08 meV and ~2.19 ± 0.63 meV obtained for X$^-$, XX$^-$ and X$^{2-}$, respectively (Supplementary Fig. S4).

The origin of the PL doublet in Fig. 2f from X$^{2-}$ can be strongly corroborated by the fact that the PL intensity of the lower energy peak is about twice that of the higher energy one, which can be similarly found in other single CsPbI$_3$ NCs (see Supplementary Figs. S2 and S3). This fixed PL intensity ratio of X$^{2-}$ has been well explained in previous optical studies of single epitaxial QDs involving FSSs in the initial and final recombination states[30,37,43] (see the schematic in Supplementary Fig. S5 and related discussions). The FSS of X$^{2-}$, together with those of X and XX$^-$, can provide valuable information on the degrees of various e-h EIs in perovskite CsPbI$_3$ NCs. The FSS of XX$^-$ is equal to the isotropic 1e$^1$-1h$^1$ EI energy ($\Delta_0^1$) and that of X to the anisotropic 1e$^1$-1h$^1$ EI energy ($\Delta_1^1$)[36,44]. The FSS of X$^{2-}$ gives the isotropic 2e$^1$-1h$^1$ EI energy ($\Delta_0^2$), while its anisotropic 2e$^1$-1h$^1$ EI energies ($\Delta_1^2$ and $\Delta_2^2$) can be generally neglected (see refs. 30 and 37, as well as the Supplementary Information for further discussions). In the above notations (1e$^1$-1h$^1$ and 2e$^1$-1h$^1$), the number before and the superscript of e (h) denote the shell and the number of electrons (holes) participating in the EI interaction. As listed in Table I for five representative CsPbI$_3$ NCs, the average values of $\Delta_1^1$ and $\Delta_0^2$ are 345 ± 110 μeV and 344 ± 28 μeV, respectively. The average value of $\Delta_0^1$ is ~290 ± 50 μeV, which is smaller than those of epitaxial CdSe/ZnSe (~1.6 ± 0.1 meV)[36,44] and InGaAs (~0.6 meV)[37] QDs, leading to a relatively larger ratio of ~1.27 ± 0.58 for $\Delta_1^1/\Delta_0^1$ in CsPbI$_3$ NCs. In contrast, the $\Delta_1^1/\Delta_0^1$ ratio is ~0.5 for CdSe/ZnSe QDs[36] and $\Delta_1^1 \ll \Delta_0^1$ in the case of InGaAs QDs[37], signifying again the important role played by lattice asymmetry[30] in the FSS of perovskite $CsPbI_3$ NCs.

To summarize, we have reported FSSs in various exciton species of single perovskite $CsPbI_3$ NCs, from which their isotropic and anisotropic e-h EI energies can be roughly estimated to favour an overall understanding of the underlying electronic processes. Although a large number of single-photon emitters have been discovered in artificially fabricated/synthesized fluorescent materials over all these years[28], it is regretful to see that only epitaxial QDs are equipped with atomic-like features in terms of their exciton FSSs and the associated coherent optical properties. However, the near-infrared PL of epitaxial QDs makes it inefficient to investigate the fundamental properties of these FSS states as well as to manipulate them in practical applications. Now that the perovskite $CsPbI_3$ NCs have stepped into the "artificial atom" regime with their bright-exciton FSSs at the visible-wavelength range, we are confident that great research interests would be stimulated in this novel type of colloidal semiconductor nanostructure to facilitate its potential usages in spin, entangled photon-pair and quantum-information-processing applications.

**Acknowledgements**

This work is supported by the National Natural Science Foundation of China (Nos. 11574147, 91321105, 11274161 and 11321063), Jiangsu Provincial Funds for Distinguished Young Scientists (No. BK20130012), and the PAPD of Jiangsu Higher Education Institutions.

**Author contributions**

X.W., C.Z., and M.X. conceived and designed the experiments. C.S., W.Y. and Y.Z. prepared the samples. C.Y., L.C., Y.L. and F.H. performed the optical experiments. C.Y. and X.W. analyzed the data. X.W., C.Y., C.Z. and M.X. co-wrote the manuscript.

**Additional information**

The authors declare no competing financial interests. Correspondence and requests for materials should be addressed to X.W., Y.Z. or M.X.

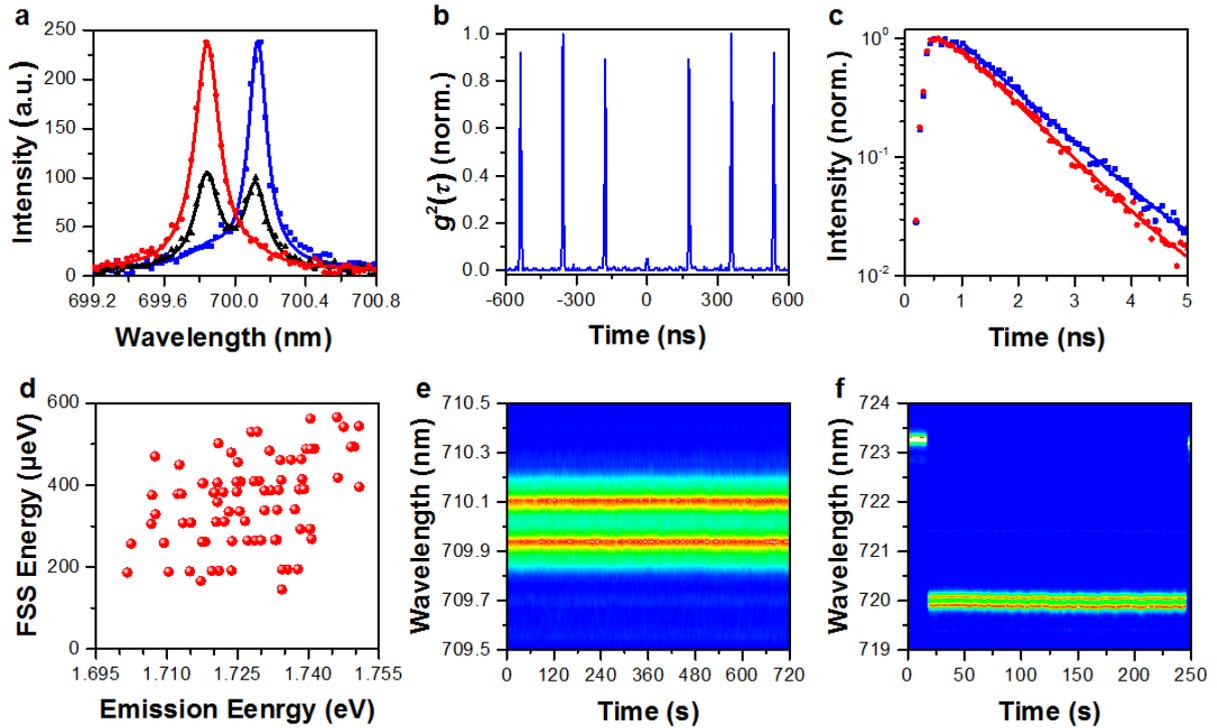

**Figure 1 | Single-exciton FSSs in single CsPbI$_3$ NCs. a**, PL spectra of a single CsPbI$_3$ NC measured at the polarization angles of 0 ° (red), 45 ° (black), and 90 ° (blue), respectively. **b**, Second-order photon correlation measurement for this single CsPbI$_3$ NC at the polarization angle of 45 °. **c**, PL decay curves measured for the higher (red) and lower (blue) energy peaks of this single CsPbI$_3$ NC and fitted with single-exponential lifetimes of ~0.93 ns and ~1.02 ns, respectively. **d**, Statistical distribution of the FSS energies plotted for ~80 single CsPbI$_3$ NCs as a function of their emission energies of the higher energy peaks. **e**, Time-dependent PL spectral image of a single CsPbI$_3$ NC showing the long-time stability of the FSS. **f**, Time-dependent PL spectral image of a single CsPbI$_3$ NC with the appearance of both X and X$^-$. The laser excitation power for the optical measurements in a-e corresponded to $N$ = ~0.05, while that in f corresponds to $N$ = ~0.8. The integration time for acquiring the PL spectra in a (e and f) was 5 s (1 s). The PL measurements in e and f were performed without any polarization selection.

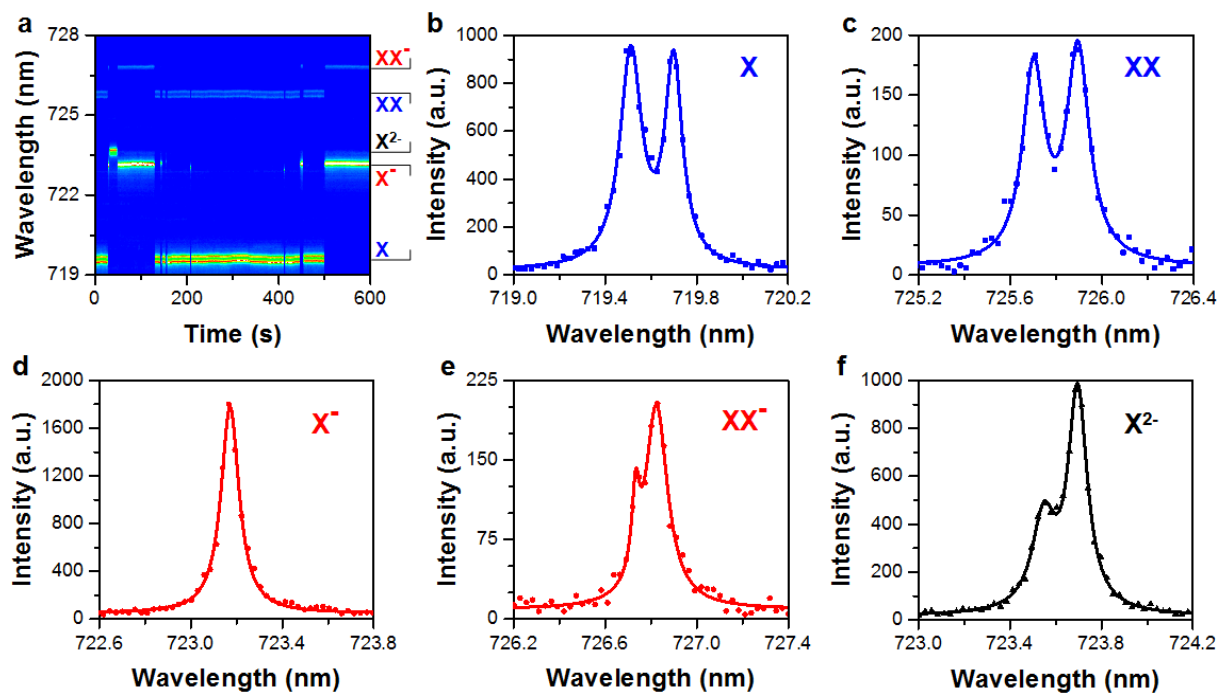

**Figure 2 | PL spectral properties of various exciton species in a single CsPbI$_3$ NC. a**, Time-dependent PL spectral image of a single CsPbI$_3$ NC excited at $N$ = ~1.5 with the appearance of various exciton species marked on the right by XX$^-$, XX, X$^{2-}$, X$^-$ and X, respectively. The PL spectra of X, XX, X$^-$, XX$^-$, and X$^{2-}$ are plotted in **b**, **c**, **d**, **e** and **f**, respectively. The integration time for all the above PL measurements is 1 s. All of the above PL measurements were performed without any polarization selection.

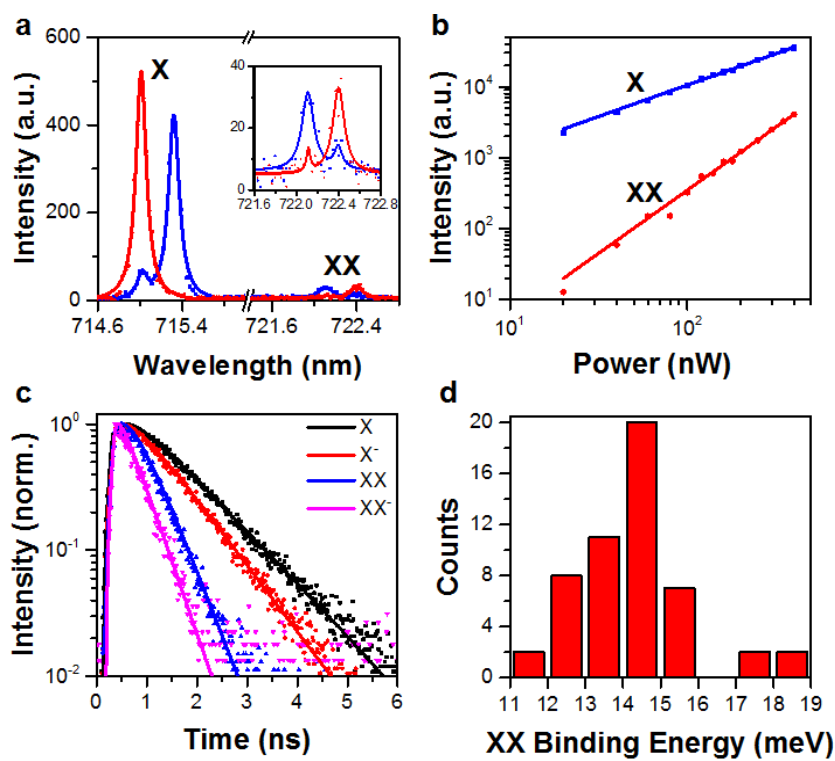

**Figure 3 | Optical properties of XX in a single CsPbI$_3$ NC. a**, PL spectra measured at the polarization angles of 0 °(red) and 90 °(blue) for X and XX of a single CsPbI$_3$ NC excited at $N$ = ~1.5. Inset: An enlarged view of the XX PL spectra. **b**, PL intensities of X and XX of this single CsPbI$_3$ NC plotted as a function of the laser excitation power with linear fitting slopes of ~0.89 and ~1.78, respectively. **c**, PL decay curves obtained at $N$ = ~1.5 for X, X$^-$, XX and XX$^-$ of this single CsPbI$_3$ NC and fitted with single-exponential lifetimes of ~1.07 ns, ~0.83 ns, ~0.41 ns and ~0.36 ns, respectively. **d**, Statistical distribution of the XX binding energies measured for ~50 single CsPbI$_3$ NCs with an average value of ~14.26 ± 1.53 meV. The XX binding energy was calculated between the average energies of the X and the XX PL doublets. The integration time for the PL measurements in a and b is 1 s.

| NC No. | $\Delta_0^2$ | $\Delta_0^1$ | $\Delta_1^1$ | $\Delta_1^1 / \Delta_0^1$ |
|---|---|---|---|---|
| I | 350 | 214 | 448 | 2.093 |
| II | 304 | 319 | 372 | 1.166 |
| III | 348 | 282 | 442 | 1.567 |
| IV | 337 | 273 | 264 | 0.967 |
| V | 382 | 356 | 201 | 0.565 |

**Table 1 | e-h EI energies of five representative CsPbI$_3$ NCs.** $\Delta_0^2$, $\Delta_0^1$ and $\Delta_1^1$ denote respectively the isotropic 2e$^1$-1h$^1$, the isotropic 1e$^1$-1h$^1$ and the anisotropic 1e$^1$-1h$^1$ EI energies in the unit of μeV.

**Bright-Exciton Fine Structure Splittings in Single Perovskite Nanocrystals**

Chunyang Yin,[1] Liyang Chen,[1] Yan Lv,[1] Fengrui Hu,[1] Chun Sun,[2] William W. Yu,[2] Chunfeng Zhang,[1] Xiaoyong Wang,[1*] Yu Zhang,[2*] and Min Xiao[1,3*]

[1]National Laboratory of Solid State Microstructures, School of Physics, and Collaborative Innovation Center of Advanced Microstructures, Nanjing University, Nanjing 210093, China

[2]State Key Laboratory on Integrated Optoelectronics and College of Electronic Science and Engineering, Jilin University, Changchun 130012, China

[3]Department of Physics, University of Arkansas, Fayetteville, Arkansas 72701, United States

[*]Correspondence to X.W. (wxiaoyong@nju.edu.cn), Y.Z. (yuzhang@jlu.edu.cn), and M.X. (mxiao@uark.edu).

**Experimental Section**

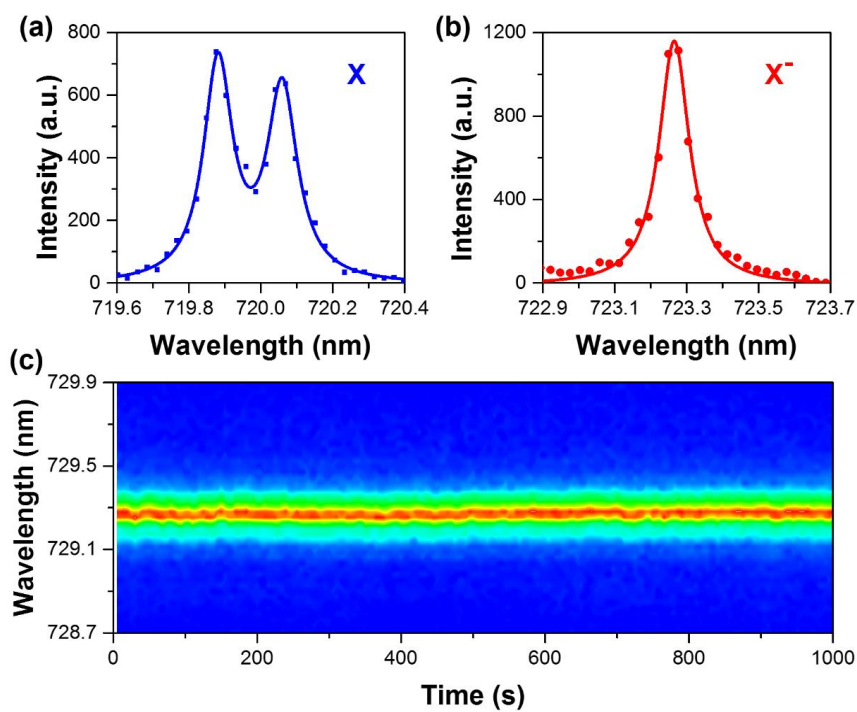

**Figure S1.** PL spectra of (a) X and (b) X⁻ of a single CsPbI$_3$ NC excited at $N = $ ~0.8. (c) Time-dependent PL spectral image of X⁻ in a single CsPbI$_3$ NC excited at $N = $ ~0.1. The PL integration times for (a), (b) and (c) were 1 s, 1 s, and 5 s, respectively. All of the above PL measurements were performed without any polarization selection.

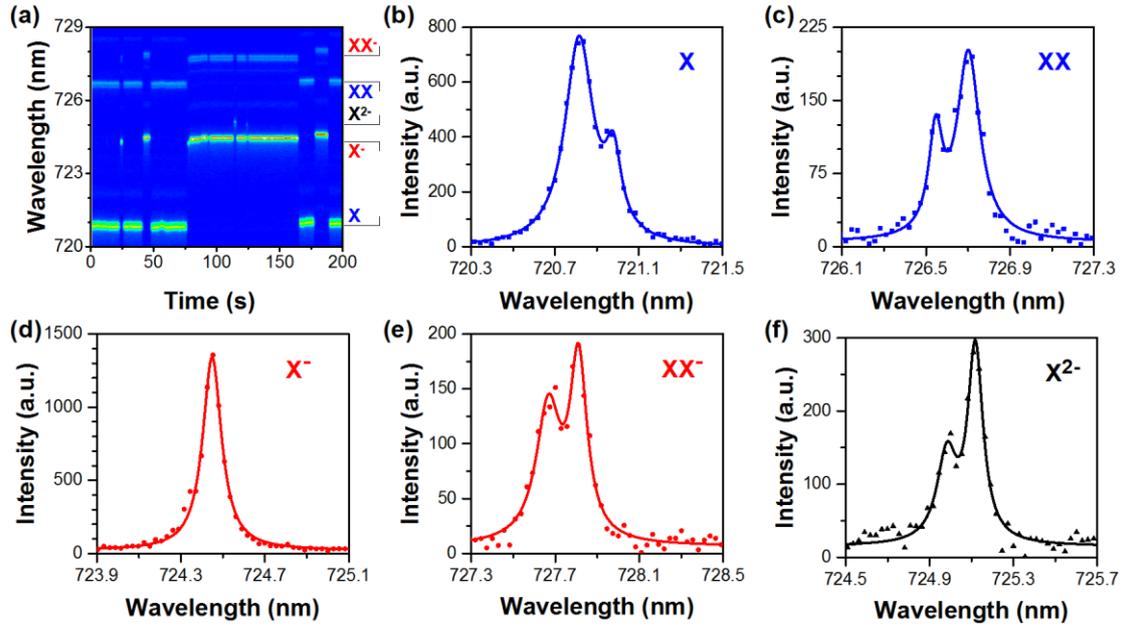

**Figure S2.** (a) Time-dependent PL spectral image of a single CsPbI$_3$ NC excited at $N$ = ~1.5 with the appearance of various exciton species marked on the right by XX$^-$, XX, X$^{2-}$, X$^-$ and X, respectively. The PL spectra of X, XX, X$^-$, XX$^-$, and X$^{2-}$ are plotted in (b), (c), (d), (e) and (f), respectively. The integration time for all the above PL measurements was 1 s. All of the above PL measurements were performed without any polarization selection.

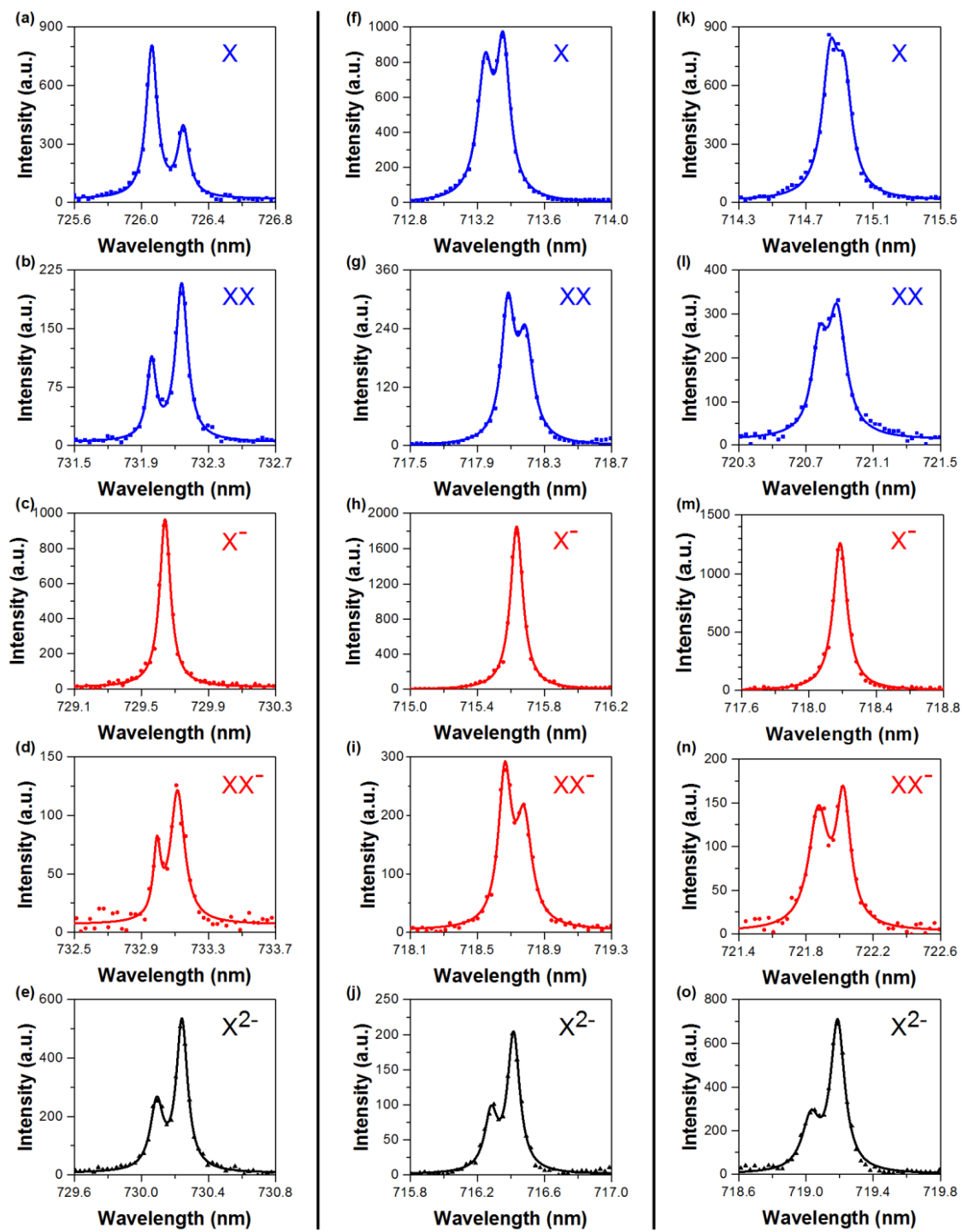

**Figure S3.** PL spectra of X, XX, X$^-$, XX$^-$, and X$^{2-}$ plotted in (a)-(e), (f)-(j) and (k)-(o) for three different CsPbI$_3$ NCs excited at $N$ = ~1.5 with an integration time of 1 s. All of the above PL spectra were collected without any polarization selection.

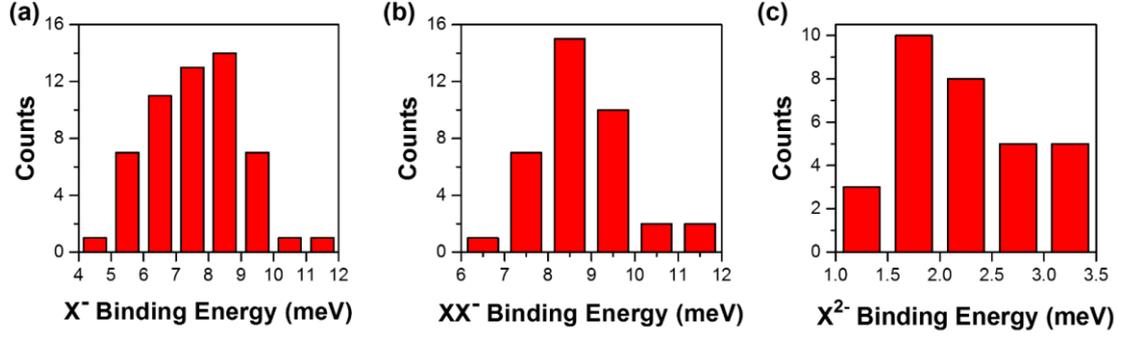

**Figure S4.** Statistical distributions of the (a) X⁻, (b) XX⁻ and (c) $X^{2-}$ binding energies obtained from 55, 37 and 31 single CsPbI₃ NCs, respectively. The X⁻ binding energy was calculated between the average energy of the X PL doublet and the energy of the X⁻ PL peak. The XX⁻ binding energy was calculated between the average energy of the XX⁻ PL doublet and the energy of the X⁻ PL peak. The $X^{2-}$ binding energy was calculated between the average energy of the $X^{2-}$ PL doublet and the energy of the X⁻ PL peak.

**Discussion Section:** Energy level structure of $X^{2-}$

From the theory of invariants (see refs. 30 and 37 in the main text), the initial state ($2e^1$-$1e^2$-$1h^1$) of $X^{2-}$ consists of one electron in the second electron shell (2e), two electrons with antiparallel spins in the first electron shell (1e) and one hole in the first hole shell (1h) with a pseudospin of either $\pm\frac{1}{2}$ or $\pm\frac{3}{2}$. The most probable transition occurs between an electron in 1e and a hole in 1h and, thus, the final state ($2e^1$-$1e^1$) is left with one electron in 2e and one electron in 1e.

If the hole pseudospin is $\pm\frac{1}{2}$ (Fig. S5a), the total spin of the initial state can be described by $|\pm 1\rangle = |\pm\frac{1}{2}\rangle|\pm\frac{1}{2}\rangle$ and $|\pm 0\rangle = |\pm\frac{1}{2}\rangle|\mp\frac{1}{2}\rangle$, where the

former $|\pm\frac{1}{2}>$ denotes the hole pseudospin and the latter $|\pm\frac{1}{2}>$ or $|\mp\frac{1}{2}>$ denotes the electron spin. The four states are split by the isotropic $2e^1$-$1h^1$ EI ($\Delta_0^2$) and the anisotropic $2e^1$-$1h^1$ EI ($\Delta_1^2$ and $\Delta_2^2$). Normally, the splitting between the $|+1>$ and the $|-1>$ states is labeled $\Delta_1^2$, the splitting between the $|+0>$ and the $|-0>$ states is labeled $\Delta_2^2$, while the splitting between the average energy of the $|+1>$ and the $|-1>$ states and the average energy of the $|+0>$ and the $|-0>$ states is labeled $\Delta_0^2$. The anisotropic $2e^1$-$1h^1$ EI energies of $\Delta_1^2$ and $\Delta_2^2$ are generally small so that the splitting between the $|+1>$ and the $|-1>$ states, as well as between the $|+0>$ and the $|-0>$ states, can be neglected. The two electrons in the final state have two kinds of configurations, one with antiparallel spins forming a singlet ($|0,0>$) and the other with parallel spins forming a triplet ($|1,+1>$, $|1,0>$ and $|1,-1>$). In the above notations, the singlet and the triplet are labeled $|S, S_z>$, where $S$ represents the total spin angular momentum and $S_z$ the projection of $S$. The singlet and the triplet are split by the electron-electron EI, which is almost one order of magnitude larger than the e-h EI since one electron and one hole are distinguishable while two electrons are not. Moreover, the splittings in the triplet are much smaller than $\Delta_0^2$ and the neglected $\Delta_1^2$ and $\Delta_2^2$ so that the three triplet states are nearly degenerate. The possible transitions of $X^{2-}$ from the initial to the final states are schematically presented in Figure S5a, from which an intensity ratio of 1:2 is expected from the initial states of $|\pm1>$ and $|\pm0>$ to their respective final triplet states.

If the hole pseudospin is $\pm\frac{3}{2}$ (Fig. S5b), the total spin of the initial state can be described by $|\pm 2> = |\pm\frac{3}{2}>|\pm\frac{1}{2}>$ and $|\pm 1> = |\pm\frac{3}{2}>|\mp\frac{1}{2}>$, where the four states are split by $\Delta_0^2$, $\Delta_1^2$ and $\Delta_2^2$ as well. Without any hole, the final states are just the same as those described earlier in the $\pm\frac{1}{2}$ case of the hole pseudospin. The possible transitions of $X^{2-}$ from the initial to the final states are schematically shown in Figure S5b, from which an intensity ratio of 1:2 is expected from the initial states of $|\pm 1>$ and $|\pm 2>$ to their respective final triplet states.

In our experiments, the transitions from the initial state of $|\pm 1>$ correspond to the higher energy peak (see Figure 2f in the main text, Figure S2f, as well as Fig. S3e, j and o) because of the lower intensity. Finer splittings ($\Delta_1^2$ and $\Delta_2^2$) in the higher and lower energy peaks of $X^{2-}$ is hard to be determined based on our current system resolution.

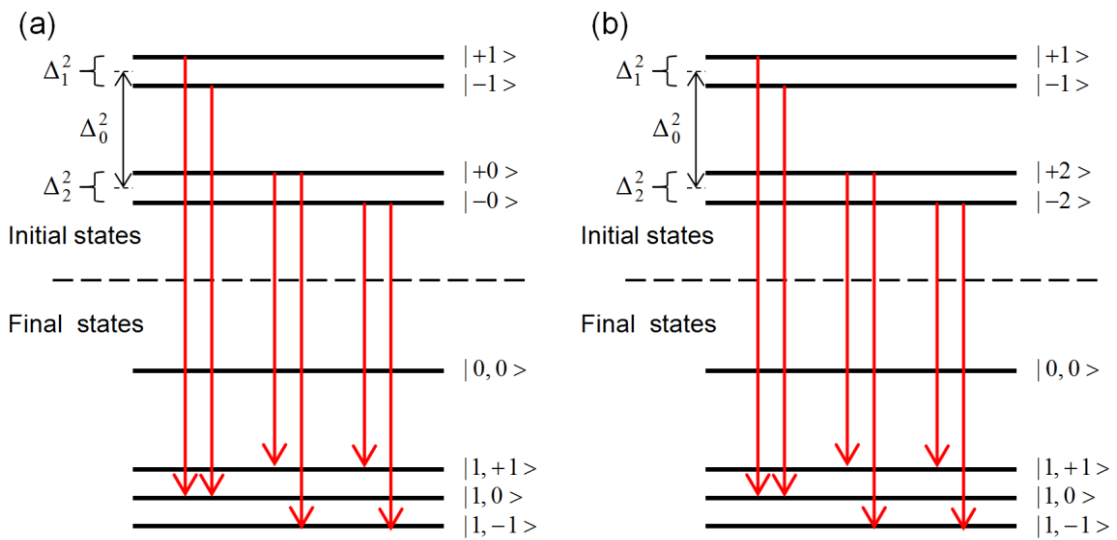

**Figure S5.** Schematics showing the $X^{2-}$ initial and final states with hole pseudospins of (a) $\pm\frac{1}{2}$ and (b) $\pm\frac{3}{2}$, respectively.